\documentclass[useAMS,usenatbib,letters]{mn2e}
\usepackage{times}

\usepackage{graphicx}

\title[Mergers and the Fundamental Plane]{From the Tully-Fisher relation to 
the  Fundamental Plane through Mergers}
\author[Aceves \& Vel\'azquez]{H\'ector Aceves\thanks{E-mail:
aceves@astrosen.unam.mx} and H\'ector Vel\'azquez \\
Instituto de Astronom\'{\i}a, Universidad Nacional Aut\'onoma de 
M\'exico, Apdo. Postal 877, Ensenada,  BC 22800, M\'exico}
\begin{document}

\date{Accepted ---- . Received ------; in 
original form --------}

\pagerange{\pageref{firstpage}--\pageref{lastpage}} \pubyear{2005}
\maketitle
\label{firstpage}

\begin{abstract}
We set up a series of self-consistent $N$-body simulations to investigate 
the fundamental plane of merger remnants of spiral galaxies. These last ones 
are obtained from a theoretical Tully-Fisher relation at $z$=1, assuming 
a constant mass-to-light ratio within the $\Lambda$CDM 
cosmogony. Using a S\'ersic growth curve 
and an orthogonal fitting method, we found that the fundamental 
plane of our merger remnants is described by the relation 
$R_\mathrm{e}\! \propto \!\sigma_0^{1.48\pm 0.01} I_{\mathrm e}^{-0.75 \pm 0.01}$ which is  
in good agreement with that reported from the Sloan Digital Sky Survey 
$R_{\mathrm e}\! \propto \! \sigma_0^{1.49 \pm 0.05} I_{\mathrm e}^{-0.75\pm 0.01}$. However, 
the $R^{1/4}$-profile leads to a fundamental plane 
given by $R_\mathrm{e}\! \propto \! \sigma_0^{1.79\pm 0.01} I_{\mathrm e}^{-0.60 \pm 0.01}$. 
In general, the correlation found in our merger remnants arises from homology 
breaking ($V^2 \propto \sigma_0^\nu$, 
$R_g\propto R_\mathrm{e}^\eta$) in combination with a 
mass scaling relation between the total and luminous mass, 
$M\propto M_\mathrm{L}^{\gamma}$. Considering an orthogonal fitting method, 
it is found that $1.74 \la \nu \la 1.79$, $0.21 \la \eta \la 0.52$ 
and $0.80 \la \gamma \la 0.90$ depending on the 
adopted profile (S\'ersic or $R^{1/4}$). 

\end{abstract}



\section{Introduction}

The formation and evolution of elliptical galaxies continues to be an  
outstanding problem in astrophysics (e.g. Meza et al. 2003, Wright et
al. 2003). In the current CDM models of hierarchical structure formation, 
mergers are the common process through which larger objects form from 
small ones (Frenk et al. 1988, Kauffmann \& White 1993, Lacey \& Cole 1993). 
In this scenario, ellipticals are a `natural' outcome (Toomre 1977) and  
accumulating observational evidence seems to support this 
(Schweizer 1998 and references therein). However, an important challenge 
to the merger hypothesis is to explain how this seemingly random merging 
process can lead to the tight correlations found in ellipticals (White 1997).

One of this correlations found in ellipticals is the so-called Fundamental 
Plane (FP); a linear relation, in the logarithmic space, among their 
effective radius, $R_{\mathrm e}$, their effective surface 
brightness, $I_{\mathrm e}$, and their one-dimensional central  
velocity dispersion, $\sigma_0$ \citep{dd87,Dressler87}:  $R_\mathrm{e} 
\!\propto \! \sigma_0^a I_\mathrm{e}^b$. The FP exhibits a 
general trend that follows from the application of the virial theorem to 
ellipticals: $M \propto V^2 R_\mathrm{g}$, 
with $M$ being the total mass, $V^2$ the 3-dimensional mean square 
velocity and $R_\mathrm{g}$ the gravitational radius of the system 
($V^2=2K/M$ and $R_\mathrm{g}=G M^2/|W|$, where $K$ and $W$ are 
the total kinetic 
and potential energy, respectively). Assuming that ellipticals are 
homologous systems in equilibrium 
($ V^2 \! \propto \! \sigma_0^2$, $R_\mathrm{g} \! \propto \! 
R_\mathrm{e}$) with a constant mass-to-light ratio, 
the virial theorem leads to
\begin{equation}\label{eq:virial}
R_\mathrm{e} \propto \sigma_0^2 I_\mathrm{e}^{-1}\, \to \,
\log R_\mathrm{e} \approx 2[\log \sigma_0 + 0.2  \mu_\mathrm{e}] 
\end{equation}
where $\mu_\mathrm{e}=-2.5 \log I_\mathrm{e}$. Deviations of an observed 
FP from the theoretical expectation are attributed to variations of 
the mass-to-light ratio with 
luminosity and/or to the breaking of 
homology in galaxies (e.g. Trujillo, Burkert \& Bell 2004).

The observed FP presents some differences depending on the 
wave-band spectrum and on the environment where galaxies reside 
\citep{jfk96,Pahre98,mobasher,Kelson2000,Bernardi03}.  Recently, 
from a sample of about $9000$ elliptical galaxies taken from the 
Sloan Digital Sky Survey (SDSS), Bernardi et al. (2003) obtained a FP 
relation given by $R_\mathrm{e}\! \propto \! \sigma_0^{1.49 \pm 0.05} 
I_\mathrm{e}^{-0.75\pm0.01}$; where the $R^{1/4}$--profile was assumed.

On the theoretical side, simulations of spiral mergers are numerous 
and have been extensively explored to study: 
the morphology of merger remnants (e.g. Mihos \& Hernquist 1996),
the line-of-sight velocity distribution (e.g. Bendo 
\& Barnes 2000), and the effect of the orientation of the angular momentum 
of discs (e.g. Naab \& Burkert 2003). Other works have considered the
effect of mergers on the properties of the FP (e.g. Capelato et al. 1995, 
Levine \& Aguilar 1996, Bekki 1998, 
Dantas et al. 2003, Nipoti et al. 2003, G\'onzalez-Garc\'{\i}a 
\& van Albada 2003). A common feature of previous studies is that they 
have used models with properties chosen rather arbitrary resembling 
present day spiral galaxies. In particular, those works addressing the FP 
of merger remnants have not considered self-consistent models of disc galaxies.

Disc galaxies also define a tight correlation between the maximum of their 
rotation velocity  $V_\mathrm{m}$ and their luminosity, termed the
 Tully-Fisher 
(TF) relation, that can be written as
$L=\mathcal{A} V_\mathrm{m}^{\alpha}$; 
where $\alpha$ and $\mathcal{A}$ are the `slope' and the zero-point, 
respectively. 
The observed values for $\alpha$ lie between $\approx\,$2.5-4, depending 
on the waveband (e.g. Strauss \& Willick 1995). In particular, the 
$I$-band TF relation is given by 
\begin{equation}\label{eq:tfrel}
M_{I} - 5 \log h = -21.00 -7.68(\log W -2.5)
\end{equation} 
where $M_I$ is the absolute magnitude of the disc, and
$W\approx 2\,V_\mathrm{m}$ is the 21-cm hydrogen line width resulting in 
a slope of $\alpha \approx 3.1$ (Giovanelli et al. 1997). 

In this {\it Letter} we study whether merger remnants of self-consistent
$N$-body spirals satisfying the TF relation can lead to the FP of ellipticals.
Spirals properties are obtained from the Press-Schechter 
formalism (Press \& Schechter 1974) in combination with the model of 
disc galaxy formation of Mo, Mao \& White (1998, hereafter MMW) and the 
TF relation under the $\Lambda$CDM model. 
The rest of this work has been structured 
as follows: in $\S$\ref{sec:model} 
we summarise the model used to set up the theoretical TF 
relation, the initial conditions and the orbital 
parameters for our simulations. In $\S$\ref{sec:results} the results and a 
discussion are given.

\section{Numerical Model}\label{sec:model}
In this section we summarise the method and properties of our models to 
study the FP associated to merger remnants.

\subsection{Tully-Fisher relation of spirals}

To construct a theoretical TF we use the Press-Schechter framework of 
hierarchical clustering and the model of MMW for the 
formation of galaxy discs. In this model five 
parameters are required to obtain the radial scale-length of a disc formed 
inside a spherical dark halo. These are the circular velocity, $V_\mathrm{c}$, 
the spin parameter, $\lambda$, the concentration of the halo, $c$, 
the fraction of disc to halo mass, $m_\mathrm{d}$  and the 
fraction of angular momentum in the disc, $j_\mathrm{d}$. To obtain
 $(V_\mathrm{c}, 
\lambda, c, m_\mathrm{d}, j_\mathrm{d})$ we have proceeded as in 
Shen, Mo \& Shu (2002). The $\Lambda$CDM cosmology was adopted to obtain 
the particular properties of disc galaxies 
(with a mass density parameter of $\Omega_0$=0.3, a cosmological constant 
of $\Omega_\Lambda$=0.7, the Hubble's constant $h$=0.7, and the perturbation 
power-spectrum normalization $\sigma_8$=1).

We have used the theoretical TF at $z$=1, an epoch thought to be close 
to the one where formation of the discs occurred (Peebles 1993), 
as our fiducial redshift to construct the numerical disc galaxies. 
This allowed us to consider remnants resulting from binary encounters and 
to compare them with ellipticals that might have formed through these 
encounters. Present day massive ellipticals are probably the result of 
more than a binary merger, but their study is out of the scope of this work.

We have selected only disc galaxies with a stability criterion 
$\varepsilon_\mathrm{m}$$\,\ge\,$$0.9$, where 
$\varepsilon_\mathrm{m}$=$V_\mathrm{m} 
(G M_\mathrm{d} / R_\mathrm{d})^{-1/2}$, and $V_\mathrm{m}$ is the maximum
 rotation velocity (Efstathiou, Lake \& Negroponte 1982; Syer, Mao \& Mo 1997).
  Galaxies were evolved in isolation during 2 Gyr, about twice 
    the time needed for the first pericentric passage in the 
    encounter. Despite the fact that our galaxy models match the previous 
    criterion some form a well defined bar (B) while in other 
    cases they develop a transient bar (T) or remain stable (S) 
    (see last column of Table~\ref{tab:glxmodels}).

\begin{figure}
\centering
\includegraphics[width=8.5cm]{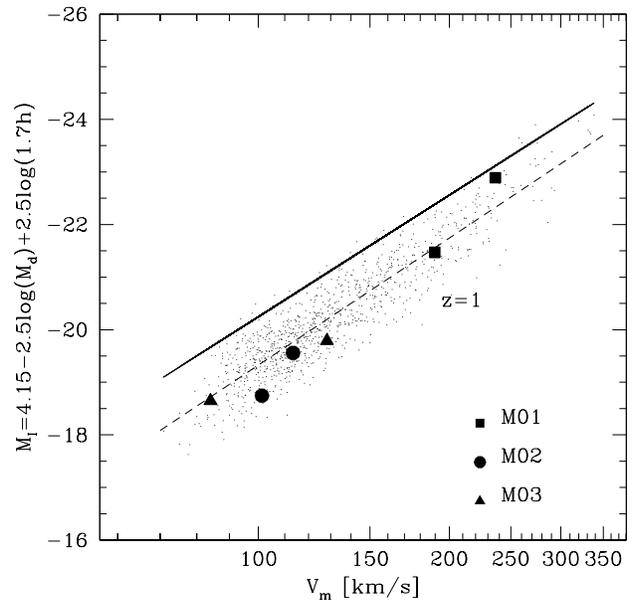}
\vspace{-0.7cm}
\caption{Theoretical Tully-Fisher relation for an ensemble of 
spirals (dots) at $z$=1. The dashed and solid lines correspond to the 
linear-square fit with a slope of $\alpha=-8.0$ and to the observed 
relation (2), respectively. The properties of discs of galaxies to be 
merged were selected from this ensemble. As an illustration, progenitors of 
mergers $M01$, $M02$ and $M03$ of Table~1 are shown.}
\label{fig:tfzetas}

\end{figure}

 In Figure~\ref{fig:tfzetas} the TF relation, 
at redshift $z$=$1$, is shown.
The straight solid line indicates the observed slope of the TF in
 the $I$-band of 
equation (2) at the present epoch. A sample of twenty random points from this
TF relation were selected in order to set up the self-consistent 
$N$-body disc galaxies.

\subsection{Galaxy models}\label{sec:galmodels}

\begin{table*}
 \centering
 \begin{minipage}{7in} 
  \caption{Properties of merging galaxies.}\label{tab:glxmodels}
  \begin{tabular}{l|crccrr|ccccrrr}
  \hline
   Merger     &  \multicolumn{6}{c}{Halo Properties} & \multicolumn{7}{c}{Disc Properties}           \\
  & $M_\mathrm{h}$ & $r_\mathrm{200}$ & $\lambda$ & $j$
   & $c$ & $N_{\rm h}$ &$f_\mathrm{d}$ & $R_\mathrm{d}$ & $z_\mathrm{d}$ & $Q$ & $R_\mathrm{Q}$ & $N_{\rm d}$  &  \\
& [M$_\odot$] & [kpc] &  &  & & &  & [kpc] & [kpc] & & [kpc] & &\\
 \hline
 $M01$  & $4.02\times 10^{11}$ & 104.3 & 0.063 & 0.041 & 7.63 & 57126 & 0.0498 & 2.4 & 0.39 & 0.402 & 5.8 &  12000 & T\\
      & $1.25\times 10^{12}$ & 152.3 & 0.056 & 0.068 & 3.84 &  177571 & 0.0586 & 5.7 & 0.47 & 0.321 & 13.8 & 44267 & T\\
\hline
 $M02$  & $1.39\times 10^{11}$ & 73.3 & 0.021 & 0.052 & 8.94 & 172403 & 0.0312 & 1.2 & 0.12 & 0.651 & 3.0 & 57436  & T \\
      & $6.45\times 10^{10}$ & 56.7 & 0.028 & 0.033 & 6.61 &  80000 & 0.0236 & 1.3 & 0.13 & 1.302 & 3.1 & 20000  & S \\
\hline
 $M03$  & $8.65\times 10^{10}$ & 62.6 & 0.078 & 0.033 & 11.12 & 149138 & 0.0402 & 1.7 & 0.20 & 0.995 & 4.1 & 42093  & S \\
      & $4.64\times 10^{10}$ & 50.8 & 0.031 & 0.043 & 12.38 & 80000 & 0.0357 & 0.7 & 0.12 & 0.789 & 1.8 & 20000 & B \\
\hline
 $M04$  & $2.36\times 10^{11}$ & 87.4 & 0.036 & 0.029 & 9.17 & 80000 & 0.0393 & 1.1 & 0.19 & 0.387 & 2.6 & 20000  & T \\
      & $1.75\times 10^{11}$ & 79.2 & 0.048 & 0.054 & 5.29 & 59428 & 0.0372 & 3.3 & 0.58 & 1.064 & 8.0 & 15671  & S \\
\hline
 $M05$  & $5.42\times 10^{10}$ & 53.5 & 0.053 & 0.114 & 15.95 & 74829 & 0.0817 & 1.2 & 0.19 & 0.561 & 2.8 & 20000  & B  \\
      & $6.02\times 10^{10}$ & 55.4 & 0.034 & 0.077 & 12.05 & 83124 & 0.0541 & 1.0 & 0.14 & 0.610 & 2.3 & 14245 & S  \\
\hline
 $M06$  & $7.49\times 10^{10}$ & 59.6 & 0.098 & 0.059 & 7.91 & 91546 & 0.0638 & 2.5 & 0.35 & 0.943 & 6.1 & 25000 & B \\
      & $6.68\times 10^{10}$ & 57.4 & 0.078 & 0.043 & 13.32 & 81742 & 0.0420 & 1.9 & 0.23 & 1.253 & 4.5 & 14327 & S \\
\hline
 $M07$  & $4.88\times 10^{10}$ & 51.7 & 0.074 & 0.064 & 11.02 & 73643 & 0.0635 & 1.5 & 0.25 & 0.841 & 3.6 & 20000 & B \\
      & $9.48\times 10^{10}$ & 64.5 & 0.047 & 0.103 & 11.79 & 143062 & 0.0656 & 1.6 & 0.31 & 0.579 & 3.9 & 42119 & B  \\
\hline
 $M08$  & $9.77\times 10^{10}$ & 65.2 & 0.032 & 0.034 & 6.56  & 245462 & 0.0239 & 1.7 & 0.20 & 1.261 & 4.1 & 36000  & S \\
      & $1.02\times 10^{11}$ & 66.1 & 0.023 & 0.012 & 7.80  & 188124 & 0.0127 & 0.9 & 0.12 & 1.217 & 2.2 & 19725 & S \\
\hline
 $M09$ & $8.11\times 10^{10}$ & 61.2 & 0.099 & 0.142 & 10.87 & 150000 & 0.0818 & 4.6 & 0.73 & 1.217 & 11.1 & 30000 & S \\
  & $8.33\times 10^{10}$ & 61.8 & 0.122 & 0.095 & 10.01 & 154050 & 0.0791 & 3.9 & 0.56 & 1.089 & 9.6 & 29686 & S\\
\hline
 $M10$ & $4.74\times 10^{11}$ & 110.3 & 0.112 & 0.106 & 11.39 & 240000 & 0.0807 & 6.6 & 1.29 & 0.576 & 15.9 & 60000 & B \\
  & $6.99\times 10^{10}$ & 58.3 & 0.071 & 0.035 & 9.94 &165992 & 0.0396 & 1.6 & 0.19 & 1.092 & 3.9 & 48689  & S \\
\hline
\end{tabular}
\end{minipage} 
\end{table*}

Our $N$-body disc galaxy models consist of a disc and a spherical dark halo 
component. Following Hernquist (1993), disc particle positions are 
randomly drawn from 
the axisymmetric density profile:
\begin{equation}
\rho_\mathrm{d} (R,z) = \frac{M_\mathrm{d}}{4 \pi R^2_\mathrm{d} 
z_\mathrm{d}} \exp( - 
R/R_\mathrm{d})  \, \mathrm{sech}^{2} (z/z_\mathrm{d}) \;,
\end{equation}
being $R_\mathrm{d}$ and $z_\mathrm{d}$ the radial and vertical scale 
length of the 
disc, respectively; $z_\mathrm{d}$ was chosen randomly  in the range 
$(0.1-0.2)R_\mathrm{d}$. For the halo we have adopted a modified version of 
the NFW--profile (Navarro, Frenk \& White 1997) with an exponential 
cut-off given by:
\begin{equation}
\rho_\mathrm{h} (r) = \frac{M_\mathrm{h}\, \alpha_\mathrm{h} }{4 \pi r  
(r + r_\mathrm{s})^2} 
\ \exp\left[ - \left(\frac{r}{r_\mathrm{200}} + q \right)^2  \right]  \;,
\end{equation}
with
$$
\alpha_\mathrm{h} = \frac{\exp( q^2) }{ \sqrt{\pi} q \exp(q^2) 
\mathrm{Erfc}(q) + 
\frac{1}{2} \exp(q^2) \mathrm{E}_\mathrm{1}(q^2) - 1 } \,,
$$
being $\mathrm{Erfc}(x)$ the complementary error function and 
$\mathrm{E}_\mathrm{1}(x)$ the 
exponential integral, $r_\mathrm{s}$ and $r_\mathrm{200}$ the scale and 
virial radii of 
the halo respectively, 
$c \!= \! q^{-1}\!= \!(r_\mathrm{s}/r_\mathrm{200})^{-1}$ the 
halo  concentration, and $M_\mathrm{h}$ is the halo total mass. Finally, 
velocities are derived from Jeans equations.

The parameters characterizing our disc galaxy models are listed in 
Table~\ref{tab:glxmodels}; where $f_{\rm d}\!=\!M_{\rm d}/(M_{\rm h}+M_{\rm d})$. We have also included the radius $R_Q$ at which 
Toomre's $Q$ parameter is normalized and the number of particles in the halo, $N_{\rm h}$, and the disc, $N_{\rm d}$. Notice the range of concentrations of 
haloes and galaxy disc sizes.

Our disc galaxy models do not include a central bulge component since, in the 
framework of MMW's galaxy formation, they tend to produce smaller disc sizes 
which require a better spatial resolution and smaller timesteps and, hence, 
more demanding computational resources.  It is not clear how this central component will affect the FP of the remnants; it will be addressed in a future work.

\subsection{Initial conditions and experiments}

We have considered only parabolic encounters where the initial separation 
between galaxy centres is 
$
R_\mathrm{s} = 1.25  ( r_\mathrm{200,1} + r_\mathrm{200,2} ) .
$
The pericentre for each encounter is randomly selected in the interval 
$R_\mathrm{p}$$\in\,$(5,20)$\,$kpc, consistent with the energies and 
eccentricities of cosmological $N$-body simulations (Navarro, Frenk \& 
White 1995). The relative orientation between galaxy spins is also taken 
randomly.

To evolve the numerical simulations we use the parallel version of {\sc
gadget}, an N-body/SPH code with individual timesteps 
(Springel, Yoshida \& White 2001). Softening parameters of 
$\epsilon_\mathrm{d}$=35$\,$pc and $\epsilon_\mathrm{h}$=350$\,$pc were 
used for 
the disc and the halo particles, respectively. {\sc Gadget} uses a spline kernel for the softening, so that the gravitational interaction between two particles is fully Newtonian for separations larger than twice the softening parameter (Power et al. 2003).

The typical time of arrival at pericentre is about 
$1\,$Gyr, and all simulations were run for a total time of about 
$8\,$Gyr; this corresponds to about the time span 
from $z$=1 to $z$=0 in a $\Lambda$CDM scenario. Remnants are already in a 
virial state by this time. 
All of our runs were performed on a cluster consisting of 32 Pentium 
processors running at 450 MHz (Vel\'azquez \& Aguilar 2003). 
Energy conservation was better than $0.25$\% in all cases.

\section{Results and discussion}\label{sec:results}

To analyse the merger remnants a constant mass-to-light ratio, $\Upsilon$, 
is assumed; hence $I_\mathrm{e}$ is 
proportional to the effective surface luminous-mass density 
$\Sigma_\mathrm{e}$. 
We first compute their `photometric' properties ($R_\mathrm{e}, 
\Sigma_\mathrm{e}$) and then its central velocity dispersion 
$\sigma_\mathrm{0}$ inside 
an aperture of radius $R_\mathrm{e}/8$. The surface density profiles 
$\Sigma (R)$ were fitted using a S\'ersic profile \citep{Sersic68,Ciotti99,Caon93} and the  
$R^{1/4}$ profile, we also fitted the corresponding growth curves of the 
luminous part of the remnants: $M_\mathrm{L}(R)=2 \pi \int \Sigma(R)  
R \,\mathrm{d}R$.

To check how the photometric properties are affected by the adopted region to 
be fitted, we have considered an aperture with outer 
radius of $17.5\,$kpc as in Wright et al (2003) and for the inner  
radius of the fit, $\xi$, we use two different values: (1) the resolution 
of the luminous component in 
our simulations, 
$\xi \!\approx \! 2.8 \epsilon_\mathrm{d}\!\approx \!100\,$pc, 
and (2) $\xi$=300$\,$pc. This also allows us to test the effects of the 
boundaries of the fitting region on the values of the exponents $(a,b)$ of the 
FP ($R_\mathrm{e} \propto \sigma_\mathrm{0}^a \Sigma_\mathrm{e}^b$). 
Each remnant was `observed' from 100 random line-of-sights. A 
$\chi^2$--minimisation by Levenberg-Marquart method \citep{NR} was used to 
obtain $R_\mathrm{e}$ and $\Sigma_\mathrm{e}$ for each projection.
The exponents 
$(a,b)$ of the FP were fitted to the complete set of projected values of 
the remnants, 
using both orthogonal and direct fits.

\begin{figure*}
\centering
\includegraphics[width=8.4cm]{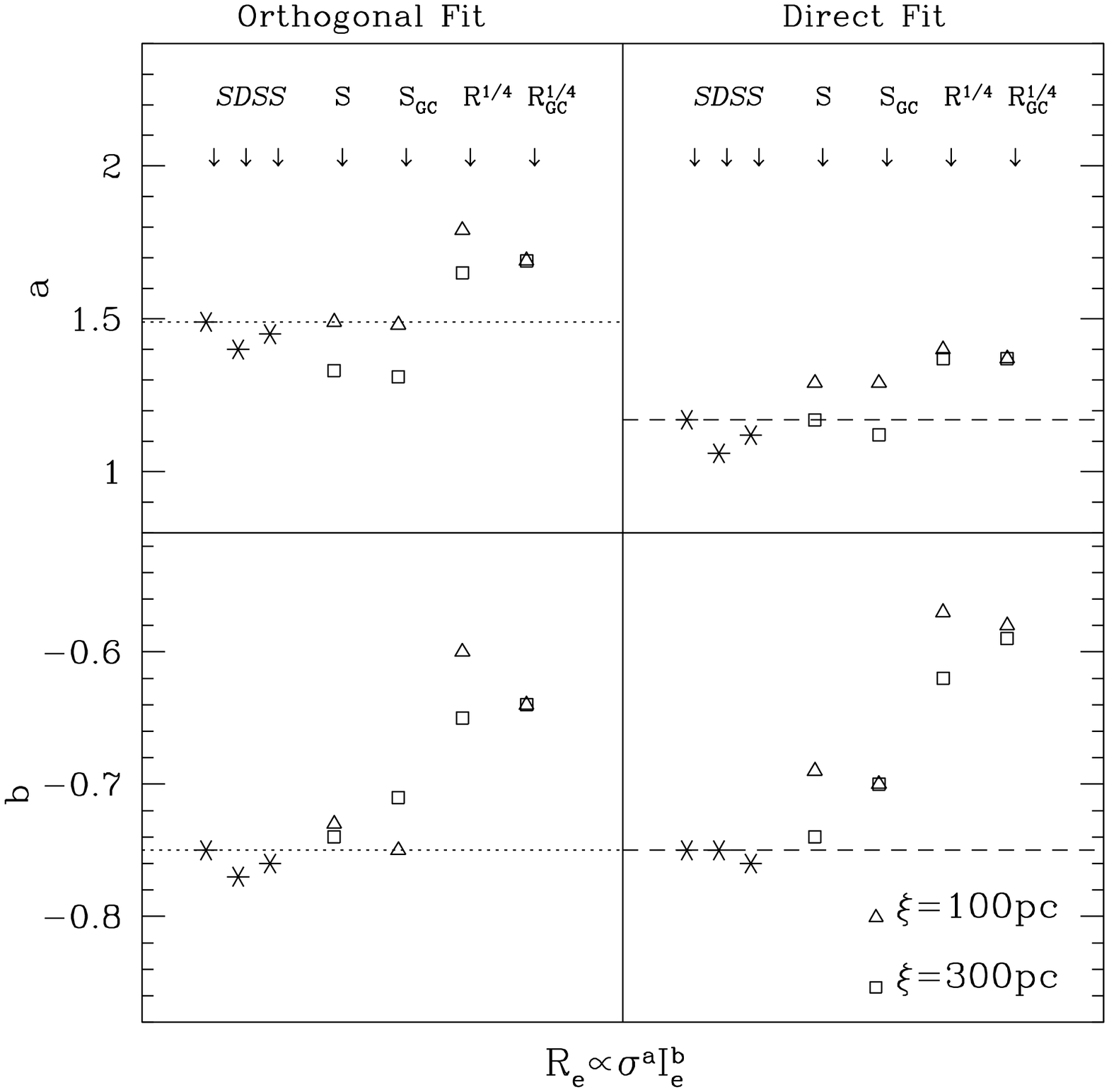}
\includegraphics[width=8.4cm]{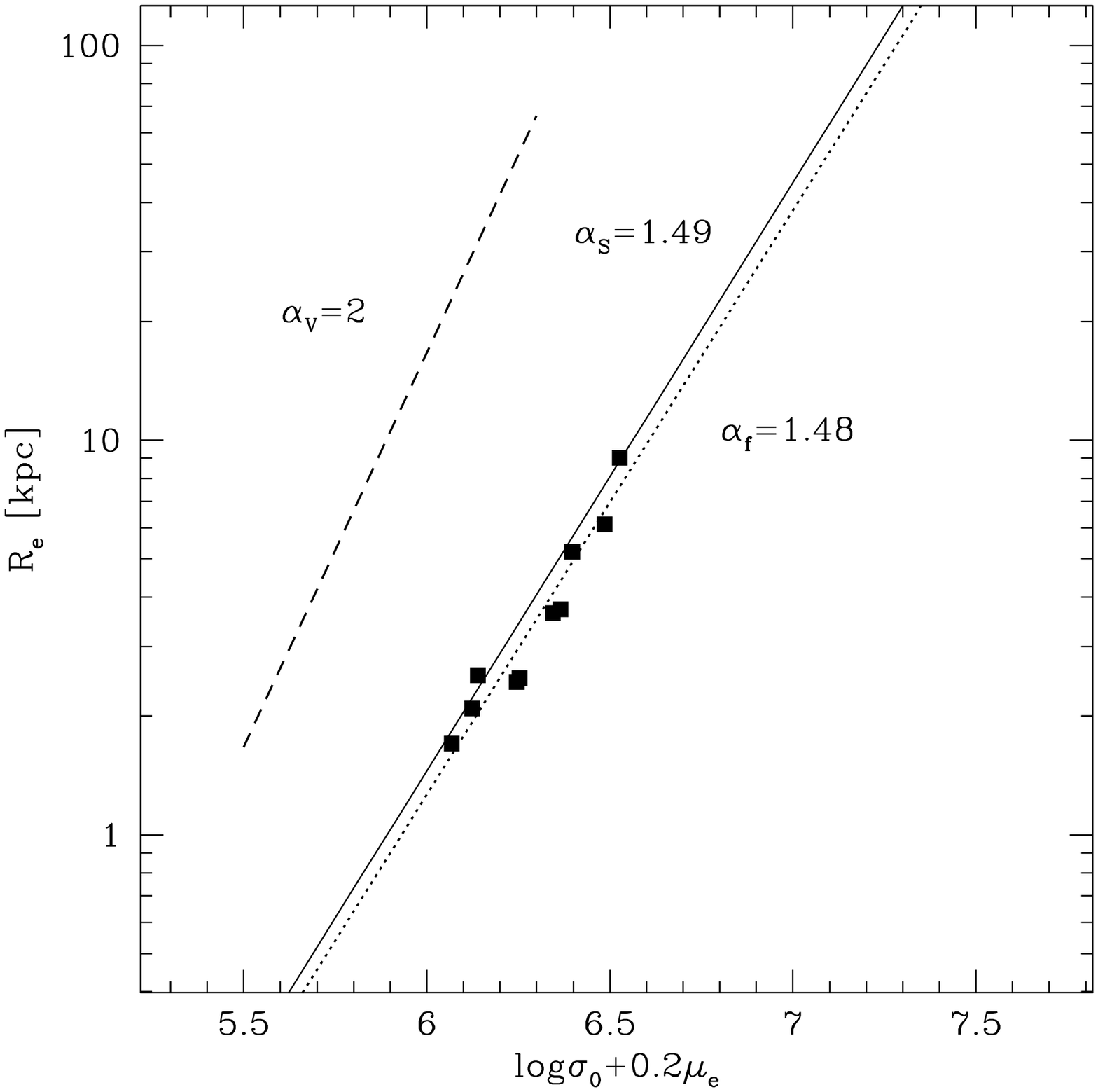}
\vspace{-0.2cm}
\caption{({\it Left}) Parameters $(a,b)$ for the FP 
($R_{\rm e} \propto \sigma_0^a I_{\rm e}^b$) of our galaxy merger remnants. 
For comparison purposes, the values of the SDSS (star symbols) 
in the $r^*$-band have been included. Horizontal dotted lines indicate the FP 
relation, $R_\mathrm{e} \propto \sigma_\mathrm{0}^{1.49} 
\Sigma_\mathrm{e}^{-0.75}$, quoted by Bernardi et al. (2003). To the left, 
these parameters were obtained using an orthogonal fitting method while to the 
right using a direct one. Here, S=S\'ersic profile, 
S$_{\rm GC}$=S\'ersic's growth curve, $R^{1/4}$=de Vaucouleurs profile, 
and $R^{1/4}_{\rm GC}$ its growth curve. Also, the effect introduced by 
two values of the inner radii $\xi$ is illustrated.  
({\it Right}) Edge-on view of the FP. Remnants are indicated by 
filled squares using the mean values listed in Table~2. 
The solid and dotted lines 
correspond to the SDSS FP ($R_\mathrm{e} \propto \sigma_\mathrm{0}^{1.49} 
\Sigma_\mathrm{e}^{-0.75}$) and to our FP$_\mathrm{f}$ model. As a reference, 
the dashed line with slope $\alpha_\mathrm{V}=2$ corresponding to 
virialised homologous systems has been plotted. Remnant mergers have been 
artificially displaced to coincide with the `zero-point' of the FP of the 
SDSS. 
}
\label{fig:fptheor}
\end{figure*}

The resulting  values for these 
exponents are plotted in Figure~\ref{fig:fptheor} ({\sl left panel}).
It can be seen how the values of parameters $(a,b)$ are 
sensitive to the fitting function (profile or growth curve), to the adopted 
profile (S\'ersic or $R^{1/4}$), to the fitting method (direct or 
orthogonal) and to the value of the radius $\xi$.  The 
observed values in the $r^*-$band of the SDSS have also been included 
(star symbols). In particular, the dotted lines refer to the exponents of 
the FP of Bernardi et al. (2003): $R_\mathrm{e} \propto
 \sigma_\mathrm{0}^{1.49} I_\mathrm{e}^{-0.75}$.

This last 
relation is quite well reproduced by the model with $\xi$=100$\,$pc, a 
S\'ersic growth curve and an orthogonal fitting method (hereafter our fiducial 
model, denoted by FP$_\mathrm{f}$) which is expressed by the relation 
$R_\mathrm{e} \propto \sigma_\mathrm{0}^{1.48\pm 0.01} \Sigma_\mathrm{e}^{-0.75 \pm 0.01}$ with a standard 
deviation orthogonal to the plane of $0.06$. Uncertainties in the exponents were estimated using the bootstrap technique \citep{Efron}.
Table~2 lists the mean projected values of the quantities used to compute our FP$_{\rm f}$ and the mean S\'ersic index.

 However, the exponents $(a,b)$ obtained by 
fitting the merger remnants with the $R^{1/4}$-profile have problems 
matching the observed ones, and this is independent of the fitting method 
(orthogonal or direct). For example, for $\xi=100\,$pc and an orthogonal 
fitting, the R$^{1/4}$-profile 
leads to a FP described by $R_\mathrm{e} \propto \sigma_\mathrm{0}^{1.79 \pm 0.01} 
\Sigma_\mathrm{e}^{-0.60\pm 0.01}$ (which will be denoted by FP$_\mathrm{dV}$), while 
for $\xi=300\,$pc it is given by 
$R_\mathrm{e} \propto \sigma_\mathrm{0}^{1.65\pm 0.01} 
\Sigma_\mathrm{e}^{-0.65\pm 0.01}$; notice that increasing $\xi$ tends to provide 
a slight better fit. An edge-view of our FP$_\mathrm{f}$ for the merger 
remnants (filled squares) 
is shown in Figure~\ref{fig:fptheor} ({\sl right panel}). As a reference, 
the long-dashed line indicates the slope, $\alpha_\mathrm{V}=2$, for 
virialised homologous systems with a constant mass-to-light ratio while 
the solid and dotted lines refer to the 
SDSS and to our FP$_\mathrm{f}$, respectively. Our merger remnants were 
artificially displaced to coincide with the zero-point of the 
FP of the SDSS (Bernardi et al. 2003).

\begin{table}
 \centering
 \begin{minipage}{8cm} 
  \caption{Mean of projected values used to determine our FP$_{\rm f}$.}\label{tab:remnants}
  \begin{tabular}{lccrc}
  \hline
   Merger     &  \multicolumn{4}{c}{Parameters$^*$}        \\
 &  $\langle n \rangle $ & $\langle  R_\mathrm{e}\rangle $ & $ \langle \sigma_\mathrm{0}\rangle $ &  $\langle  \mu_\mathrm{e}\rangle $ \\
 &  & [kpc] & [km/s] & \\
 \hline
 $M01$  & 4.3  & 9.0 & 139 & -19.4 \\
 $M02$  & 2.2 & 2.5 & 75 & -19.4\\
 $M03$  & 1.8 & 2.2 & 72 & -19.4 \\
 $M04$  & 2.6 & 3.7 & 95  & -19.4 \\
 $M05$  & 3.8  & 1.7 & 71 &-20.3\\
 $M06$  & 2.3 & 3.7 & 67 &-18.8  \\
 $M07$  & 3.2 & 2.5& 66& -19.8  \\
 $M08$  & 2.8 & 2.1 & 45 & -19.1 \\
 $M09$  & 2.1 & 6.1 & 72 & -18.3 \\
 $M10$  & 2.6 & 5.2& 71& -18.7  \\
\hline
  \multicolumn{5}{l}{$^*\,$  Here $n$ is the index of the S\'ersic profile [$\propto \exp(-R^{1/n})$], } \\
  \multicolumn{5}{l}{ and $\mu_\mathrm{e}=-2.5\log \Sigma_\mathrm{e}$ with $\Sigma_\mathrm{e}$ in M$_\odot/$kpc$^2$.}  \\
\end{tabular}
\end{minipage} 
\end{table}

The overall agreement of the parameter values $(a,b)$ for the S\'ersic profile 
case with those of the SDSS relies on  a combination of homology breaking 
($V^2 \propto \sigma_0^\nu$, $R_g\propto R_\mathrm{e}^\eta$) and 
 a mass scaling relation between the total and luminous mass as 
$M\propto M_\mathrm{L}^{\gamma}$. For instance, for our 
FP$_\mathrm{f}$ non-homology is found to be 
$V^2 \propto \sigma_\mathrm{0}^{1.74}$ and
$R_{\rm g} \propto R_\mathrm{e}^{0.51}$, 
with a mass scaling relation given by $M\propto M_\mathrm{L}^{0.85}$ and  
a negligible dependence of S\'ersic's index with the effective radius 
($n \propto R_\mathrm{e}^{0.07}$),
 while  for the FP$_\mathrm{dV}$ case, 
these relations are described by $V^2 \propto \sigma_\mathrm{0}^{1.77}$, 
$R_\mathrm{g} \propto R_\mathrm{e}^{0.21}$ and $M\propto M_\mathrm{L}^{0.80}$. 
Note that a direct fitting method shows larger differences between 
our fitted FP and that of the SDSS (see Figure~\ref{fig:fptheor}-{\sl right panel}).  Despite that S\'ersic's profile fitting leads to a better match with observations; a consistent comparison will require an observational FP determined using this type of profile on the large data base of the SDSS.

 The above results suggest that, ignoring any contribution from 
breaking-homology 
relations, it does not appear feasible to scale the mass-to-light ratio, 
$(M/L)$, as a simple power-law of the luminosity, $L$, (or equivalently of 
the mass $M$) alone; see also Bernardi et al. (2003). 
 Perhaps, a more general relation of the form 
$(M/L) \propto R_\mathrm{e}^x \sigma_0^y L_{\rm e}^z$ need to be explored. 
A more extensive study of the scaling relations of merger 
remnants, their non-homology, the distribution of luminous and dark matter, 
and  their kinematics is under way and it will be presented in a future work.

Our results support the idea that mergers of spirals, with properties 
obtained from the
 Tully-Fisher relation within a $\Lambda$CDM cosmogony, lead to the tight 
correlation expressed by the FP.

 In dissipational numerical simulations of mergers, it is found that a gas component can yield important differences in the stellar component of the remnants in comparison with their collisionless counterparts (e. g., Barnes \& Hernquist 1996). For instance, a gas component deepens the potential well, affecting more the remnant kinematics than its luminous profile. However, it is not clear from these results how the FP is going to be altered; in particular, if its slope is going to be changed or just the dispersion of values around it. This remains an open issue.

\section*{Acknowledgments}

We thank Luis Aguilar and Simon White for their helpful comments on this work. 
This research was funded by CONACyT-M\'exico Project 37506-E. An anonymous referee is thanked for useful comments that helped to clarify some points of this work.



\bsp
\label{lastpage}
\end{document}